\documentclass[10pt,journal,compsoc]{IEEEtran}

\usepackage{multirow}
\usepackage{amssymb}
\usepackage{amsmath}
\usepackage{caption}
\usepackage[caption=false,font=footnotesize,list=false]{subfig}
\usepackage{tabularx}

\usepackage{xurl}
\usepackage{lineno}
\usepackage{graphicx}
\usepackage{hyperref}          

\hypersetup{
    breaklinks=true,
    colorlinks=true,
    linkcolor=black,
    citecolor=black,
    urlcolor=blue
}
\usepackage{titlesec}
\titlespacing*{\section}      {0pt}{8pt plus 2pt minus 2pt}{4pt plus 1pt}
\titlespacing*{\subsection}   {0pt}{6pt plus 2pt minus 2pt}{3pt plus 1pt}
\titlespacing*{\subsubsection}{0pt}{4pt plus 2pt minus 2pt}{2pt plus 1pt}
\usepackage{flushend}

\begin{document}





\title{Greenness-Driven Scheduling in Far Edge Kubernetes: A CODECO Evaluation}

\author{
  \IEEEauthorblockN{Kaikang Huang$^*$, Dalal Ali, Rute C. Sofia} \\
  \IEEEauthorblockA{fortiss GmbH -- Research Institute of the Free State of Bavaria\\
  Munich, Germany\\
  \{khuang, sofia\}@fortiss.org, dalalaqlan@gmail.com}\\
  \small $^*$Corresponding author
}

\maketitle
\begin{abstract}
Energy consumption is an increasing concern in IoT-Edge–Cloud infrastructures, where containerized application orchestration must balance performance with sustainability. This
paper investigates how the Kubernetes CODECO framework integrates cross-layer energy-awareness into scheduling decisions for containerized applications across the IoT-Edge–Cloud continuum.
CODECO monitors energy at both the computational level, via Kepler, and a network (IP) level, and uses these metrics to define greenness heuristics that guide pod
placement decisions through its ILP-based scheduler.

The approach is experimentally evaluated on a real-world far Edge testbed composed of ARM-based embedded devices, comparing CODECO against vanilla Kubernetes across multiple scenarios. The results show that CODECO consistently reduces the energy consumption of the cluster, with savings of up to 11.01 \, mJ in computational energy and 4.14 \, mJ in network
transmission energy consumption at peak load, for a wide set of scenarios which combine different types of injected fault conditions, including CPU stress, asymmetric network delay, and bandwidth contention. A composite greenness score combining both energy dimensions provides a stable and consistent ranking of scheduling strategies across all conditions, demonstrating its suitability as a unified energy indicator for cluster-level orchestration decisions across the IoT-Edge–Cloud continuum.
\end{abstract}


\section{Introduction}
\label{intro}
The softwarization of network and computational resources has accelerated the decentralization of services across the Cloud-edge-IoT (CEI) continuum, pushing compute and data storage toward the far Edge, closer to a wide number of interconnected data sources. Large and complex applications, including AI models, can now be distributed across the CEI. 

Existing orchestration mechanisms such as Kubernetes support deployment decisions based on Quality of Service (QoS) requirements such as CPU and memory. However, energy consumption has yet to be treated as a first-class requirement in the orchestration
of applications across the CEI continuum. As the energy footprint of distributed computing infrastructures continues to grow, optimizing resource usage across both computational and network layers has become a pressing challenge.

To address this gap, the CODECO (Cognitive Decentralized Edge-Cloud Orchestration) framework based on Kubernetes introduces an energy-aware orchestration approach that integrates cross-layer energy metrics directly into scheduling decisions~\cite{sofia2024framework, sofia2026scalablefederatedcontainerorchestration}. By jointly considering both computational and network transmission energy, CODECO enables a more intelligent workload placement that balances performance with energy efficiency, allowing applications to operate more sustainably while maintaining the responsiveness, reliability, and scalability required, for instance, in industrial environments, as is the case with Industrial IoT (IIoT) applications.

This work presents and evaluates CODECO's cross-layer energy-aware scheduling approach considering far Edge environments. Greenness heuristics are introduced as scheduling cost functions that guide pod placement based on node-level energy profiles, and a composite greenness score is proposed as a unified indicator for cluster-level energy assessment. The approach is validated on a real far Edge testbed composed of ARM-based embedded devices, demonstrating measurable energy savings under a range of workload intensities and injected fault conditions.

This work makes three main contributions. First, it describes how CODECO monitors energy use across both computing hardware and network links, and feeds this information into Kubernetes to guide scheduling decisions. Second, it introduces and evaluates a set of "greenness" scores $g(i)$ that combine network and compute energy data into a single measure of how energy-efficient a given application placement is. Third, it tests the approach on a real Edge testbed built from ARM-based embedded devices, showing that it uses less energy than standard Kubernetes scheduling under different workload sizes and failure conditions.

This paper is organized as follows. Section~\ref{relatedwork} describes related literature, highlighting the contributions of this work. Section~\ref{codeco} introduces the CODECO framework, focusing on the components that integrate energy-awareness into the scheduling pipeline. Section~\ref{sec:codeco-observability} describes the cross-layer observability approach that collects computational and network energy metrics, and introduces the greenness heuristics and scheduling cost functions. Section~\ref{sec:experimentation-setup} presents the experimental setup, including the testbed, application, load generation, scenarios, and automation workflow.
Section~\ref{evaluation} presents the performance evaluation and analysis across proposed scenarios.
Section~\ref{conclusions} concludes the paper and proposes directions for future research.

\section{Related Work}
\label{relatedwork}
Resource scheduling in Kubernetes-based CEI infrastructures has been studied along two largely independent lines: energy-aware scheduling and network-aware scheduling. This section reviews representative work in each direction and identifies the gap that motivates the cross-layer approach presented in this paper.

\subsection{Energy-Aware Scheduling}
Energy-aware scheduling has emerged as a practical means of
addressing the growing energy footprint of distributed edge--cloud
infrastructures. Luo~\emph{et al.}~\cite{luo2019fog} propose an
energy-balancing task scheduling algorithm for container-based fog
computing that dynamically assigns tasks according to the
transmission energy states of edge nodes, balancing energy usage
among heterogeneous nodes, and prolonging system lifetime. Their work
establishes that incorporating energy metrics into scheduling
decisions can significantly improve sustainability in
resource-constrained environments, though it focuses on transmission
energy alone and does not consider computational energy or
Kubernetes-native orchestration.

More recent work has targeted power and carbon awareness directly
within Kubernetes. Kepler~\cite{kepler} provides eBPF-based
per-container power estimation by correlating hardware performance
counters with energy measurements, exposing node and
container-level power consumption through Prometheus. Although Kepler is a foundational monitoring tool — and is adopted in the present work — it does not itself influence scheduling decisions.
Carbon-aware scheduling extensions~\cite{codeco_energy} have
explored using such measurements to trigger workload migration
toward lower-carbon or lower-energy nodes, demonstrating that
energy-aware rescheduling can improve cluster stability under
high-load conditions and that composite energy models are better
suited to heterogeneous hardware than single-metric approaches.

\subsection{Network-Aware Scheduling}

Incorporating network conditions into scheduling decisions has been recognized as an effective means of improving application
performance in distributed environments.
NetMARKS~\cite{netmarks} enhances Kubernetes pod scheduling by
collecting dynamic network metrics via Istio Service Mesh, achieving
significant reductions in response time and inter-node bandwidth
consumption in 5G and edge environments. However, NetMARKS
optimizes network performance exclusively and does not consider
energy efficiency or cross-layer scheduling.

Zeus~\cite{zeus} improves resource efficiency in large-scale
Kubernetes clusters by enabling safe co-location of
latency-sensitive and best-effort workloads, highlighting the
importance of real-time resource utilization and interference
awareness during scheduling. Like NetMARKS, Zeus does not
incorporate energy-awareness into its scheduling logic.

NAS~\cite{nas} proposes an eBPF-enabled network-aware Kubernetes
scheduling framework that exploits fine-grained network metrics to
guide placement decisions. While conceptually close to the network
observability approach adopted in this work, NAS focuses exclusively
on network conditions and does not jointly consider energy
consumption, limiting its ability to address energy-network
trade-offs in heterogeneous CEI deployments.

\subsection{Research Gap and Motivation}

\begin{table}[htp!]
\centering
\caption{Comparison of related scheduling approaches.}
\label{tab:related}
\setlength{\tabcolsep}{3pt}
\renewcommand{\arraystretch}{1.1}
\begin{tabular}{|l|c|c|c|c|}
\hline
\textbf{Approach} & \textbf{E-aware} & \textbf{N-aware} &
\textbf{X-layer} & \textbf{K8s} \\
\hline
Luo~\emph{et al.}~\cite{luo2019fog} & \textbf{Y} & -- & -- & -- \\
\hline
Kepler~\cite{kepler}                 & \textbf{Y} & -- & -- & \textbf{Y} \\
\hline
NetMARKS~\cite{netmarks}             & --          & \textbf{Y} & -- & \textbf{Y} \\
\hline
Zeus~\cite{zeus}                     & --          & \textbf{Y} & -- & \textbf{Y} \\
\hline
NAS~\cite{nas}                       & --          & \textbf{Y} & -- & \textbf{Y} \\
\hline
CODECO~\cite{codeco_energy}          & \textbf{Y} & \textbf{Y} & \textbf{Y} & \textbf{Y} \\
\hline
\end{tabular}
\vspace{2pt}
\end{table}

Table~\ref{tab:related} summarizes the key properties of the reviewed approaches. Although energy-aware and network-aware scheduling have each been extensively studied, orchestration
frameworks that jointly consider both dimensions in a unified, cross-layer manner remain limited. Existing approaches typically optimise a single dimension, overlooking the strong
interdependencies between energy consumption, network conditions, and workload placement in real-world CEI systems. In particular, no existing Kubernetes-native framework combines
node-level computational energy estimation with flow-level network transmission energy in a single scheduling cost function, nor evaluates the trade-offs between these two dimensions under injected fault conditions on real embedded hardware.

This work addresses that gap by building on the CODECO framework~\cite{sofia2024framework} to introduce a cross-layer, composite approach that jointly considers computational and network transmission energy into scheduling cost functions and evaluates the resulting trade-offs experimentally on a real far Edge testbed under a range of workload intensities and fault conditions.

\section{CODECO Overview}
\label{codeco}

\begin{figure}[htp!]
\centering
\includegraphics[width=\columnwidth]{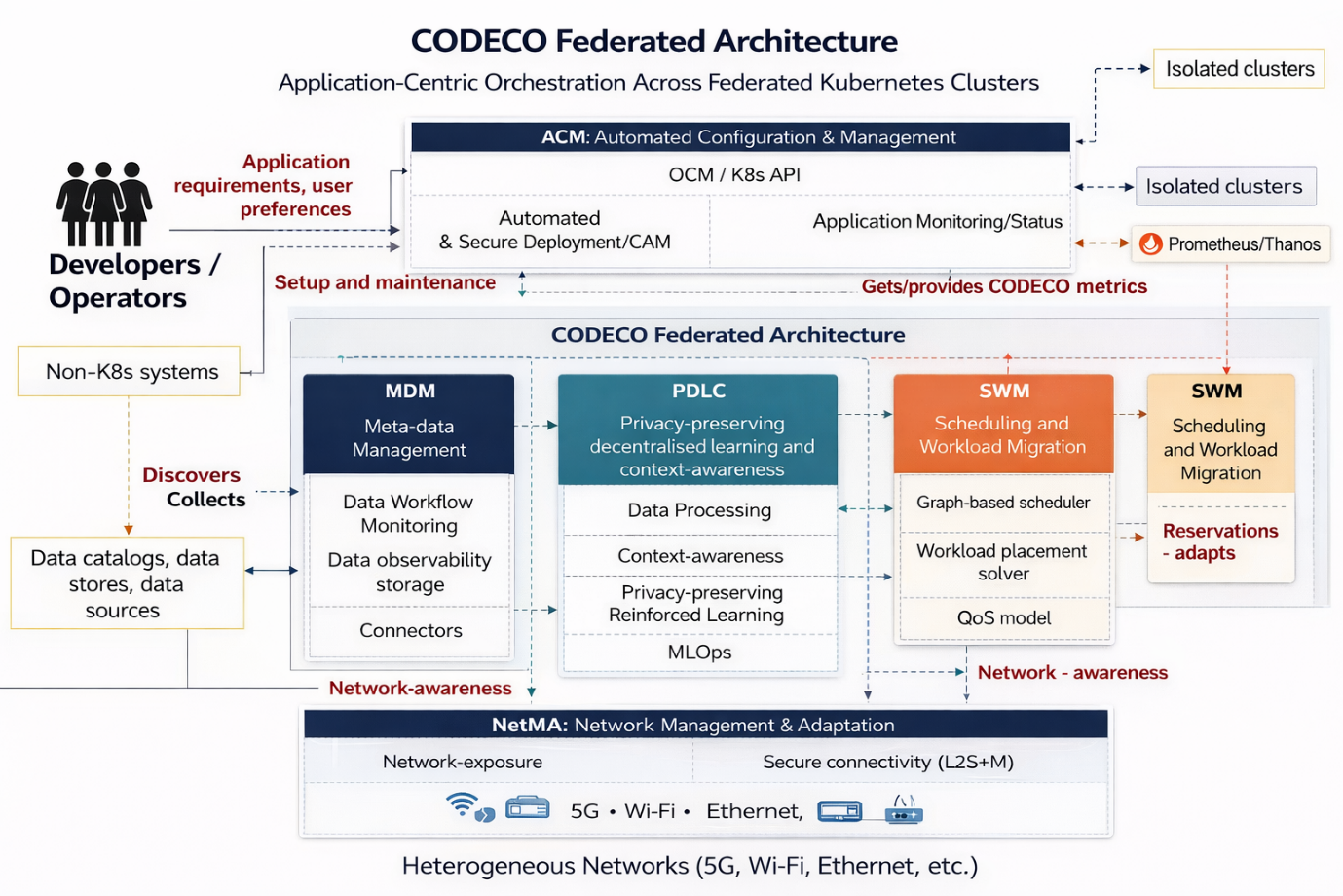}
\caption{The CODECO framework and its components.}
\label{fig:codeco}
\end{figure}

CODECO\footnote{https://gitlab.eclipse.org/eclipse-research-labs/codeco-project} is a Kubernetes-based orchestration framework designed to extend Cloud-native application orchestration capabilities toward heterogeneous and dynamic CEI environments. Its primary objective is to augment Kubernetes with cross-layer intelligence encompassing compute, network, and data dimensions, thereby enabling adaptive, context-aware, and efficient workload placement across the infrastructure continuum. The architectural overview of CODECO and its principal components is illustrated in Figure~\ref{fig:codeco}.%
\footnote{A detailed overview of the CODECO federated
orchestration capabilities is available in~\cite{CODECO-D31,sofia2026scalablefederatedcontainerorchestration}.}

At the heart of CODECO is the Advanced Configuration and Management (ACM) component, which automates both application setup and runtime adaptation. ACM is based on the RedHat Open Cluster Management (OCM)\footnote{https://open-cluster-management.io/} solution. In addition to being the single CODECO interface to the user, it incorporates computational considerations into orchestration workflows, ensuring that deployments remain aligned with the operational characteristics of the underlying infrastructure. Central to ACM is the CODECO
Application Model (CAM), a semantic description of an application's micro-services and their \textit{Quality of Service (QoS)} functional and non-functional requirements. CAM captures microservice dependencies, operational constraints, and target performance profiles such as resilience or greenness, which serve as high-level objectives guiding CODECO's orchestration decisions across all components.

The Metadata Manager (MDM) brings data workflow observability to CODECO, maintaining consistent multi-layer snapshots of the system, capturing data dependencies, device status, and infrastructure changes that influence deployment and re-deployment decisions.

The Privacy-preserving Decentralized Learning and Context-awareness (PDLC) component enriches the orchestration pipeline with context-aware intelligence and AI-driven placement
recommendations based on the performance profiles defined in CAM. It operates through two mechanisms:

\begin{itemize}
\item \textbf{Aggregated node and cluster scoring (PDLC-CA):} computes composite costs consistent with user-defined performance profiles.
\item \textbf{System Stability Estimation:} applies Multi-agent Reinforced Learning (MARL) to detect potential instability in the infrastructure and issues scheduling recommendations intended to promote robust application deployment.
\end{itemize}

Workload placement is orchestrated by the Scheduling and Workload Migration (SWM) component, which provides a graph-oriented, ILP-based Kubernetes scheduler, that considers both application requirements (CAM) and the infrastructure state monitored by CODECO to compute a near-optimal placement solution.
CODECO continuously monitors the infrastructure through ACM (compute), NetMA (network), and MDM (data), exporting all metrics to Prometheus\footnote{\url{https://prometheus.io/}} to allow seamless incorporation of new or user-defined metrics. For the purposes of this work, two perspectives are relevant:

\begin{itemize}
    \item \textbf{Computational Perspective:} node-level energy consumption, resource availability, and device constraints, collected via Kepler and exported by ACM.
    \item \textbf{Networking Perspective:} underlay and overlay performance, link energy, flow energy, and inter-node connectivity, collected by NetMA.
\end{itemize}

\section{CODECO Observability}
\label{sec:codeco-observability}
The CODECO framework incorporates a cross-layer observability approach that integrates metrics from the application, data, and network domains to support context-aware orchestration across the
CEI continuum. Measurements collected by ACM, MDM, and NetMA are exported to Prometheus for unified monitoring, analytics, and scheduling support. The observability layer serves three primary
purposes: (i)~providing real-time visibility into the state of the infrastructure, (ii)~supporting metric-driven scheduling and workload migration through SWM, and (iii)~enabling user-defined performance profiles such as energy-efficiency (\emph{CODECO greenness}) or resilience.

By integrating Kubernetes-native metrics with CODECO-specific cross-layer measurements, the observability pipeline extends traditional cluster monitoring with additional insights from MDM and NetMA, as well as context-aware processing from PDLC. This allows SWM to correlate application behavior with infrastructure conditions in a unified manner, informing runtime decisions such as pod scheduling, migration, and scaling. Of the three observability categories described in this section, computational and network observability are directly relevant to the energy-aware scheduling approach evaluated in this paper.

This section describes the three main categories of observability supported in CODECO.

\subsection{Application Observability}

Application observability in CODECO builds on the native mechanisms provided by Kubernetes, which manages the lifecycle of containers, pods, and deployments. Through ACM, CODECO collects standard
Kubernetes telemetry including pod status, readiness and liveness probes, container restart counts, resource requests and limits, and CPU/memory utilisation via the \emph{Metrics Server} and
\emph{cAdvisor}. Additional signals include event logs, scheduling delays, QoS levels, autoscaling triggers, and application-level indicators such as request latency and error rates. Together, these
metrics provide both fine-grained and aggregated visibility into application behavior and form the basis for application-level energy analysis by linking resource utilisation to computational
energy consumption.

\subsubsection{Observability of Node Energy}
Kubernetes does not natively expose power or energy metrics as part
of its resource model. Node-level energy observability instead emerges from the integration of Kubernetes with operating system interfaces and external monitoring tools. Kubernetes contributes the execution context required for energy attribution, e.g., pod and container
identifiers, \textit{cgroup} hierarchies, and per-container resource
usage from \textit{cAdvisor} and the \textit{Kubelet}, capturing CPU cycles, memory usage, I/O operations, and network activity, all of which correlate with dynamic energy consumption.

Actual energy measurements originate outside Kubernetes, through hardware and kernel interfaces such as Intel Running Average Power Limit (RAPL), Advanced Configuration and Power
Interface (ACPI), or Baseboard Management Controller (BMC) mechanisms such as Intelligent Platform Management Interface (IPMI). Frameworks such as
Kepler\footnote{\url{https://sustainable-computing.io/}} bridge this gap by combining (i)~hardware counters when available, (ii)~kernel-level instrumentation via eBPF, and
(iii)~cgroup-scoped resource statistics from Kubernetes. Kepler correlates these sources to estimate node- and container-level energy consumption and exposes the results through Prometheus. CODECO relies on the same mechanisms to observe node energy, with Kubernetes
providing the semantic structure for attribution and Kepler providing
the measurements.

In the experimental testbed described in Section~\ref{sec:testbed}, all worker nodes are ARM-based embedded devices. This introduces important observability constraints.
ARM-based devices such as Raspberry Pis typically lack the standard power telemetry interfaces available on x86 servers (ACPI, IPMI), and while ARM kernels support eBPF, the associated \emph{BPF Type Format (BTF)} provides only minimal metadata without exposing hardware-level energy counters. Direct hardware energy measurements are therefore not available on these devices.

On x86 bare-metal environments with internal sensors, Kepler can directly access kernel power data and export accurate node and container power readings. Kepler also provides pre-trained power
models derived from CPU and system-resource usage patterns~\cite{cncf2023}, but these are architecture-specific and were trained on Intel\textsuperscript{\textregistered}
Xeon\textsuperscript{\textregistered} processors. Applying them to ARM devices introduces inaccuracies due to different hardware characteristics.

The experiments presented in this paper adopt Kepler's default  x86-based power model as a practical approximation, sufficient to capture high-level energy trends for comparative evaluation while acknowledging the limitations on ARM hardware. More accurate estimation would require architecture-specific model training, as explored in~\cite{RPIs}; where retraining is not feasible, empirical approaches such as those proposed by Pol~et~al.~\cite{Pol2024}
provide an alternative for embedded platforms.

The node energy metric used in this work is collected via the following Kepler Prometheus query, which returns the cumulative dynamic energy consumed by a node over a five-minute interval,
aligned with the sampling strategy used throughout the experiments:

\begin{verbatim}
increase(kepler_node_platform_joules_total{
    exported_instance="<node.name>",
    mode="dynamic"}[5m])
\end{verbatim}

\subsection{Data Observability}
Data observability is managed through MDM, which monitors data location, lineage, freshness, access frequency, and transfer overhead between nodes and clusters, and tracks dependencies between
datasets and microservices. These metrics support data-aware orchestration by ensuring workloads are placed near required data sources. Data observability is not the focus of the present work;
the reader is referred to~\cite{sofia2026scalablefederatedcontainerorchestration} for a detailed description.

\subsection{Network Observability}
Network observability through NetMA provides visibility into connectivity and communication performance across heterogeneous networks including 5G, Wi-Fi, and Ethernet. NetMA exposes metrics
such as latency, available bandwidth, jitter, packet loss, and energy, captured from both underlay and overlay viewpoints per node and per inter-cluster link. All metrics
are exported to Prometheus, enabling the orchestration layer to avoid congested links, maintain QoS guarantees, and optimise energy-performance trade-offs. Within NetMA, the \emph{Network
State Management (NSM)} sub-component maintains an up-to-date view of the end-to-end network state across clusters.

\subsubsection{MON Sub-system and Energy Monitoring}
The netma-nsm-mon (\textit{MON}) module is a sub-component of NetMA responsible for source-node network probing and aggregation of underlay metrics. MON builds on the open-source Kubernetes \emph{k8s-netperf}\footnote{\url{https://github.com/leannetworking/k8s-netperf}} plugin, adapted and extended to support additional metrics, including energy-related ones. MON periodically monitor the parameters shown in Table~\ref{tab:netma-nsm-mon-metrics} via a cron job.

All underlay metrics are collected at a Kubernetes worker-node level, as worker nodes are the only point in the system where application execution context and end-to-end packet visibility
coexist. Network switches and routers observe only raw packets and cannot infer which pod, container, or service produced a flow. In contrast, worker nodes host the microservices and expose full visibility into their ingress and egress traffic. Metrics such as \texttt{uNodeBandWidth}, \texttt{uLatencyNanos}, \texttt{uPacketLoss}, and \texttt{uLinkEnergy} require correlating packets with the Kubernetes resources responsible for generating them, a linkage that intermediate network devices cannot provide.

Worker nodes also expose visibility across all active network interfaces, enabling monitoring of node degree, link failures, aggregated bandwidth, and link-level energy. These metrics are
collected by MON and fed into PDLC-CA, which uses them together with computational energy from Kepler to compute the $g(i)$ scheduling cost functions described in Section~\ref{sec:codeco-observability}.
\begin{table}[htp!]
\centering
\caption{Network metrics exported by \texttt{MON}.}
\label{tab:netma-nsm-mon-metrics}
\setlength{\tabcolsep}{3pt}
\renewcommand{\arraystretch}{1.4}
\footnotesize
\begin{tabular}{|l|l|l|}
\hline
\textbf{Attribute} & \textbf{Description} & \textbf{Unit} \\
\hline
\texttt{nodeName}        & Node identifier              & --   \\
\hline
\texttt{name}            & Link identifier              & --   \\
\hline
\texttt{uLinkFailure}    & Link failure count           & --   \\
\hline
\texttt{uPacketLoss}     & Packet loss rate             & --   \\
\hline
\texttt{uNodeNetFailure} & Link+endpoint failure sum    & --   \\
\hline
\texttt{uNodeBandWidth}  & Total egress bandwidth       & Mbps \\
\hline
\texttt{uNodeDegree}     & Active link count            & --   \\
\hline
\texttt{uLatencyNanos}   & Egress latency (EMA)         & ns   \\
\hline
\texttt{uLinkEnergy}     & Energy per link (EMA)        & W    \\
\hline
\end{tabular}
\end{table}
\subsubsection{Link Energy Estimation}
MON adopts the linear energy model proposed by
Feeney et al.~\cite{feeney2001}, which estimates the energy consumed by a single packet of size~$s$ (bytes) as in Equation (1):
\begin{equation}
    e(s) = \beta_1 \cdot s + \beta_0,
    \label{eq:packet_energy}
\end{equation}
where $\beta_1$ is the incremental per-byte cost and $\beta_0$ is
the fixed cost associated with protocol control overhead
(e.g.\ RTS/CTS/ACK exchange). Both coefficients depend on the
communication mode and are determined experimentally.

The link energy $l_e(i,j)$, defined as the energy consumed by
node~$i$ due to transmission across egress link~$j$ during a
monitoring period~$T$, is obtained by summing over all~$n$ packets observed on that link as provided by Equation (2):
\begin{equation}
    l_e(i,j) = \sum_{p=1}^{n} e(s_p),
    \label{eq:link_energy}
\end{equation}
where $s_p$ is the size of the $p$-th packet.
This lightweight estimation method requires no particular hardware , making it appropriate for CODECO edge deployments.

\subsubsection{MON Example}
\begin{figure}[h]
    \centering
    \includegraphics[width=\columnwidth]{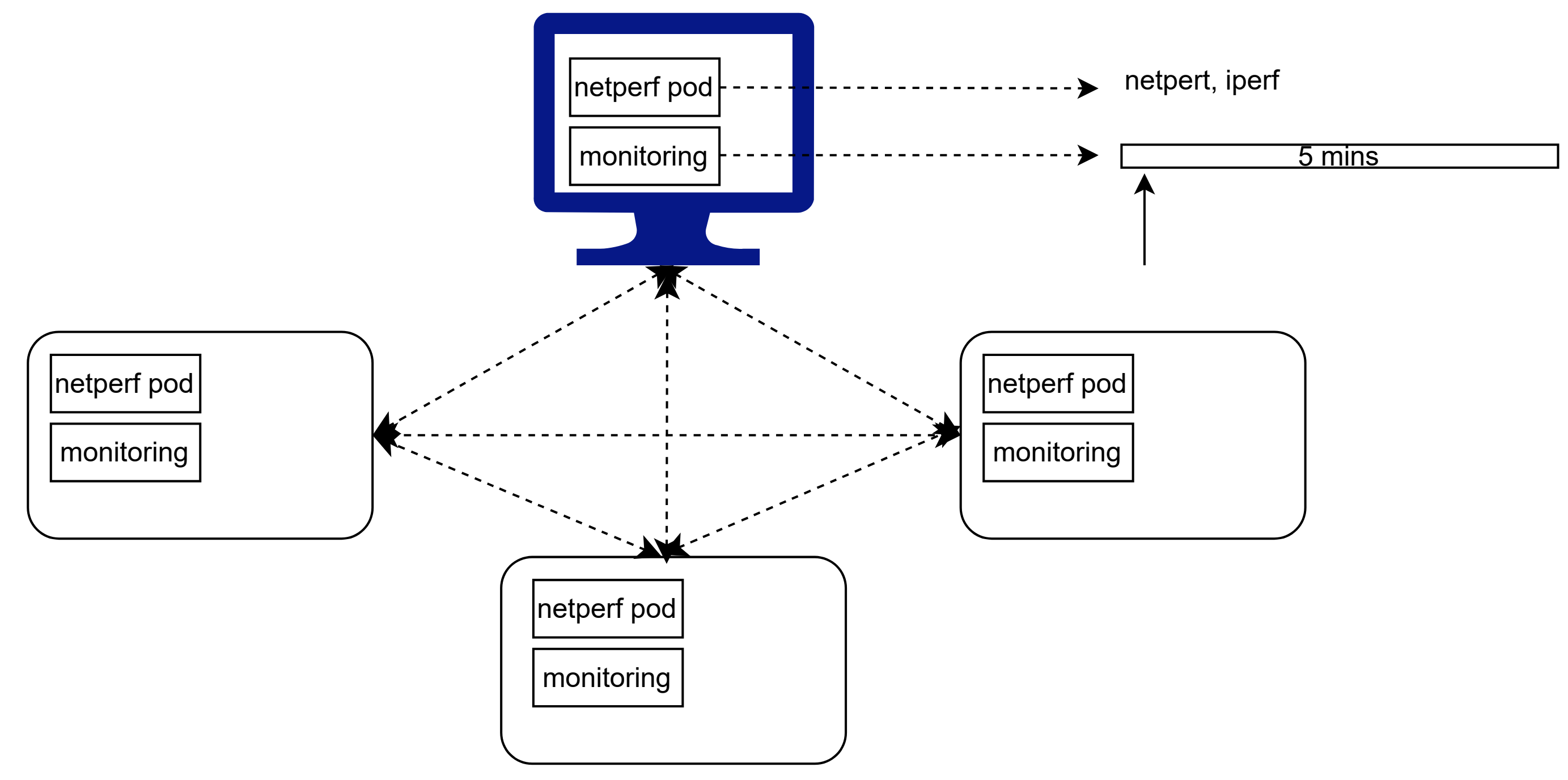}
    \caption{MON example: cluster with four nodes.}
    \label{fig:netma-example}
\end{figure}

Figure~\ref{fig:netma-example} illustrates a scenario in which a Kubernetes cluster consists of one master node and three worker nodes interconnected via Wi-Fi. Each worker hosts two daemon pods: i) a passive \texttt{k8s-netperf} pod containing the modified netperf binaries and iPerf3, and ii) an active monitoring pod responsible for executing network measurements and updating a ConfigMap. Periodic ConfigMap updates are performed (default: every five minutes), with the interval configurable by the developer.

\subsection{CODECO Greenness Metrics}

Greenness in CODECO is a flexible concept that can be aligned with
different sustainability objectives, including energy efficiency and
CO$_2$ footprinting. In this work, greenness is defined as
energy-awareness, encompassing energy consumption metrics collected
at both the computational and network levels. The following
energy-related metrics are considered:

\begin{itemize}
    \item \textbf{Node Energy ($N_e(i)$):} total energy consumed by
    node $i$ across all its active processes, collected via Kepler
    using the eBPF-based query described in
    Section~\ref{sec:codeco-observability}.

    \item \textbf{Link Energy ($l_e(i,j)$):} energy consumed by
    node $i$ due to the transmission of bits across egress link $j$,
    estimated by NetMA-MON using the linear power model of
    Equation~\ref{eq:link_energy}.


    \item \textbf{Network Energy ($L_e(i)$):} aggregate energy
    consumed across all egress links of node $i$:
    $L_e(i) = \sum_{j} l_e(i,j)$.
\end{itemize}

\subsection{Greenness Scheduling Functions}

CODECO uses scheduling cost functions $g(i)$ to rank nodes in terms
of greenness. The cost value $g(i)$ for each node is computed by
PDLC-CA and passed to the SWM scheduler to guide pod placement
decisions. Three formulations are exemplified with Equations (3,4,5):

\begin{align}
    g(i) &= N_e(i)               \label{eq:greenness_compute} \\
    g(i) &= L_e(i)               \label{eq:greenness_network}  \\
    g(i) &= N_e(i) \cdot L_e(i) \label{eq:greenness_composite}
\end{align}

Equation~\ref{eq:greenness_compute} captures a compute-centric view of
greenness, ranking nodes solely by their total computational energy
consumption. Equation~\ref{eq:greenness_network} captures a
network-centric view, ranking nodes by the aggregate transmission
energy across all egress links. Equation~\ref{eq:greenness_composite}
provides a composite view, jointly sensitive to both computational
and network transmission energy, and to the number and cost of
active links.

The choice of $g(i)$ formulation defines the greenness performance
profile applied during scheduling. The trade-offs between these
three formulations are experimentally evaluated in
Section~\ref{evaluation}.

\section{Experimental Setup}
\label{sec:experimentation-setup}
\subsection{Experimental Environment}
\label{sec:testbed}

\begin{figure}[h]
\centering
\includegraphics[width=0.8\columnwidth]{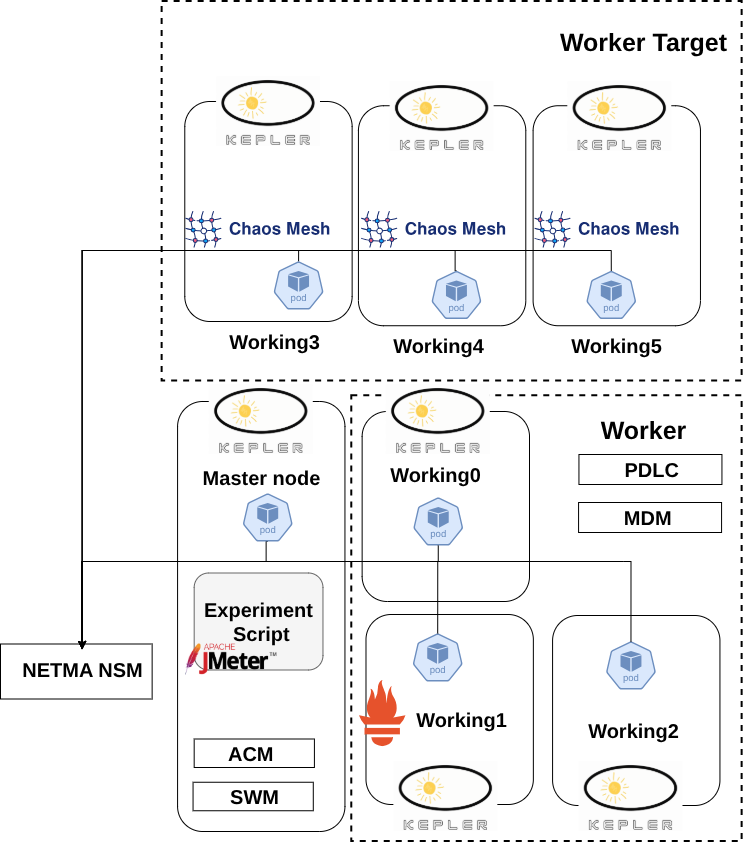}
\caption{CODECO testbed used in the experiments.}
\label{fig:experimentation}
\end{figure}

\begin{table}[htp!]
\centering
\caption{Testbed equipment features.}
\label{tab:set-up}
\setlength{\tabcolsep}{4pt}
\renewcommand{\arraystretch}{1.3}
\begin{tabular}{|l|l|l|l|}
\hline
\textbf{Role} & \textbf{Device} & \textbf{Spec.} &
\textbf{Node ID} \\
\hline
Control plane & Laptop & 16\,Gi, 8 cores & master \\
\hline
Worker        & RPI4   & 3.8\,Gi, 4 cores & working0--2 \\
\hline
Worker target & RPI4   & 3.8\,Gi, 4 cores & working3--5 \\
\hline
\end{tabular}
\end{table}
The experiments were conducted on a lightweight Kubernetes distribution, k3s\footnote{\url{https://k3s.io/}}, deployed on a live testbed hosted at the fortiss IIoT
Lab\footnote{\url{https://www.fortiss.org/en/research/fortiss-labs/detail/iiot-lab}}. Figure~\ref{fig:experimentation} illustrates the experimental setup, including the positioning of CODECO within the testbed and their interaction
with monitoring components.
The setup consists of a single k3s cluster with one master node (a laptop) and six Raspberry Pi~4 corresponding to CODECO worker nodes as summarized in Table~\ref{tab:set-up}. A distinction is made between worker nodes that are used as target for application migration, and worker nodes where the application has been first deployed. The cluster is interconnected via a private IP network based on Wi-Fi (IEEE 802.11bg). The control node runs Ubuntu with kernel version 6.8.0-57-generic, while the worker nodes use a custom-built kernel (6.6.22-codeco-v8+) integrating eBPF hooks for energy monitoring through Kepler in Raspberry PIs.

Each node runs Kepler as a Prometheus exporter. Kepler leverages eBPF to collect CPU performance counters and Linux kernel tracepoints~\cite{kepler}. During the CODECO deployment (triggered by the ACM component), both Prometheus and Kepler are launched to monitor node-level compute energy metrics. The ACM component collects compute energy data from Kepler, while the NetMA component captures network energy consumption metrics. Both metrics are continuously made available to PDLC and other CODECO components for cost evaluation (see section 3).

\subsection{Application Setup and Load Generation}
\subsubsection{Application Bookinfo}
\begin{figure}[ht!]
\centering
\includegraphics[width=0.8\columnwidth]{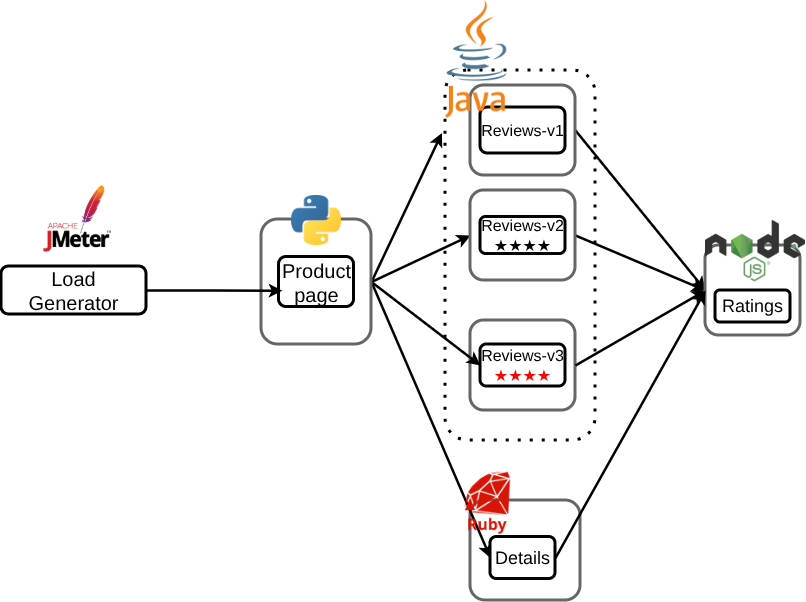}
\caption{BookInfo structure used in the experiments.}
\label{fig:bookinfo}
\end{figure}
The application used in our experimentation is Bookinfo by Istio\footnote{\url{https://istio.io/latest/docs/examples/bookinfo/}}, whose microservice topology is illustrated in Figure~\ref{fig:bookinfo}. Bookinfo comprises four micro-services:
\texttt{productpage}, \texttt{details}, \texttt{reviews}, and
\texttt{ratings}. 
In the baseline deployment, \texttt{reviews} is restricted to a single instance (\texttt{v1}), yielding four pods in total. To evaluate the impact of workload scaling on scheduling decisions, the number of \texttt{reviews} replicas is varied across experimental runs, as detailed in
Section~\ref{sec:scenario}.

Application deployment is managed through the CODECO CAM YAML file (section 3), in which developers specify all the pods (micro-services) that compose an application and their interconnections\footnote{The bookinfo CODECO yaml file used in experiments is available under \url{https://git.fortiss.org/iiot\_external/experimentation/-/tree/main/CODECO-energy-awareness-benchmarking}}.
 
\subsubsection{Load Generation with JMeter}
Apache JMeter\footnote{\url{https://jmeter.apache.org/}} issues HTTP requests to the Bookinfo \texttt{productpage} microservice at a controlled rate expressed in \textit{requests per second} (RPS), used here as a target load intensity.
Three intensity levels are evaluated: 1\,RPS (low), 15\,RPS (medium), and 30\,RPS (high).

RPS is configured via the \textit{Throughput Shaping
Timer}\footnote{\url{https://jmeter-plugins.org/wiki/ThroughputShapingTimer/}},
which paces request dispatch, and sustained by the \textit{Concurrency
Thread Group}\footnote{\url{https://jmeter-plugins.org/wiki/ConcurrencyThreadGroup/}},
which maintains the number of active threads needed to meet the target rate.
Since thread demand depends on response time, the required thread count $C$
is calculated as in Equation \ref{eq:thread-count}:

\begin{equation}
C = \frac{r \cdot t}{1000}
\label{eq:thread-count}
\end{equation}

\noindent where $r$ is the target RPS and $t$ is the expected maximum
response time in milliseconds (250\,ms), ensuring threads remain active long enough to sustain stable pressure throughout each run.

\subsection{Experimental Scenarios}
\label{sec:scenario}
Three progressive experimental scenarios are defined, each building upon the findings of the previous one: from workload characterisation under baseline conditions, through fault resilience evaluation, to a comparison of greenness scheduling strategies under combined stress.

\subsubsection{Scenario~I: Scalability and Load Characterisation}
This scenario compares CODECO against vanilla Kubernetes across a range
of workload intensities to identify the conditions under which each
approach is most applicable. Two parameters are varied independently:
the request rate (RPS) and the number of Bookinfo \texttt{reviews}
micro-service replicas.

\subsubsection{Scenario~II: Edge Fault Resilience}
\begin{figure}[h]
\centering
\includegraphics[width=\columnwidth]{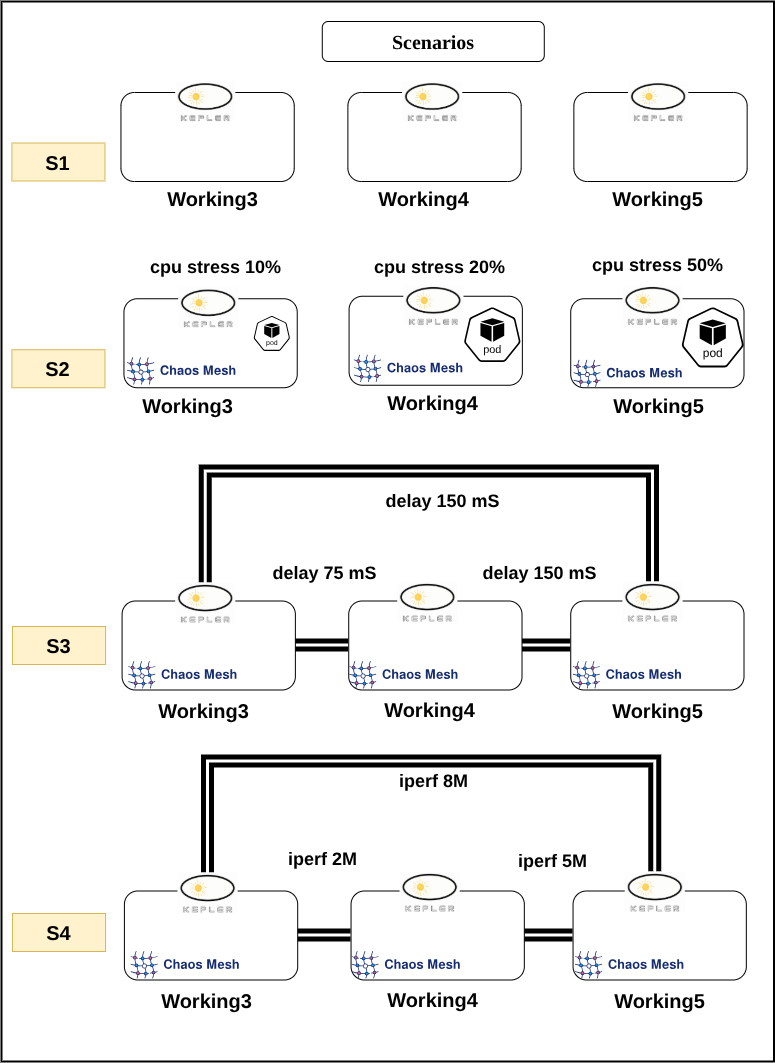}
\caption{Fault injection targets for Scenario~II: (S1)~baseline with no
         interference, (S2)~\texttt{StressChaos} CPU stress on selected
         nodes, (S3)~\texttt{NetworkChaos} latency injection on selected
         links, and (S4)~\texttt{NetworkChaos} iPerf3 generating background traffic on selected links.}
\label{fig:sub-scenarios}
\end{figure}

With the workload fixed at RPS\,=\,30 and seven \texttt{reviews} replicas, this scenario aims at creating the dynamic and unpredictable conditions typical of real-world Edge deployments. Fault injection is performed using Chaos Mesh~\footnote{\url{https://chaos-mesh.org/}}, a cloud-native chaos engineering platform that integrates with Kubernetes to inject controlled faults at the pod, node, and network level. Four sub-scenarios of increasing severity are applied, as
illustrated in Figure~\ref{fig:sub-scenarios}:
\begin{itemize}
  \item \textbf{SII.1 - Baseline:} No fault injection; both schedulers operate under nominal conditions, establishing a reference for subsequent sub-scenarios.
  \item \textbf{SII.2 - CPU Stress:}\texttt{StressChaos} faults induce heterogeneous CPU contention across the three target worker nodes: 10\% on \texttt{worker3}, 20\% on \texttt{worker4}, and 50\% on \texttt{worker5}, simulating resource-constrained edge nodes with
        varying degrees of degradation.
  \item \textbf{SII.3 - Network Delay:} \texttt{NetworkChaos} faults inject asymmetric latency across links: 75\,ms on the \texttt{worker3}--\texttt{worker4} link and 150\,ms on both the \texttt{worker3}-\texttt{worker5} and \texttt{worker4}-\texttt{worker5} links, to create conditions that could match geographically dispersed or unstable Edge connectivity.
  \item \textbf{SII.4 - Controlled bandwidth contention:} \texttt{iperf3} background traffic saturates inter-node links with asymmetric loads: 2\,Mbps on \texttt{worker3}-\texttt{worker4}, 5\,Mbps \texttt{worker4}-\texttt{worker5}, and 8\,Mbps on \texttt{worker3}-\texttt{worker5}, simulating congestion conditions typical of a loaded far Edge network.
\end{itemize}

\subsubsection{Scenario~III: Greenness Scheduling Strategies Comparison}
This scenario retains the same workload as Scenario~II (RPS\,=\,30,
seven \texttt{reviews} replicas) and fixes the interference conditions, namely, combined CPU and network stress applied via Chaos Mesh and
\texttt{iperf}, while varying the $g(i)$ scheduling function used
by PDLC-CA to implement the Greenness performance profile. The three $g(i)$ formulations provided in Equations (3,4,5) are evaluated:

\begin{itemize}
  \item \textbf{Computational (\texttt{comp}):} $g(i) = N_e(i)$; captures computational energy on the worker nodes without explicit consideration of inter-service network costs       (Equation~\ref{eq:greenness_compute}).
  \item \textbf{Network (\texttt{net}):} $g(i) = L_e(i)$; captures the energy consumption due to IP packet transmission.
  \item \textbf{Composite (\texttt{green}):} $g(i) = N_e(i) \cdot L_e(i)$; jointly captures both computational and network transmission energy consumption (Equation~\ref{eq:greenness_composite}).
\end{itemize}

The outcome of each $g(i)$ formulation is assessed across three
cluster-level score metrics: computational energy, network
energy, and the composite greenness score, which
consolidates both dimensions into a single indicator.

\begin{figure}[htp!]
\centering
\includegraphics[width=\columnwidth]{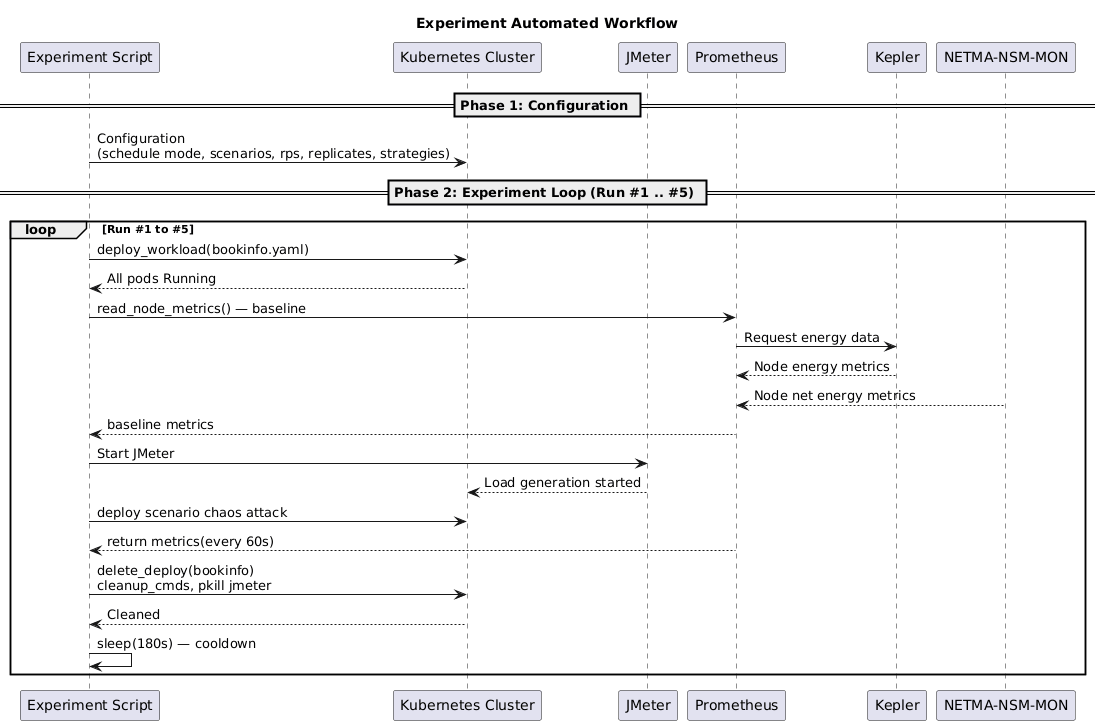}
\caption{Automated experimentation workflow.}
\label{fig:automated-workflow}
\end{figure}
\subsection{Experimental Automation Workflow}

To ensure reproducibility across configurations, all experiments were automated as detailed in the communication sequence diagram provided in Figure~\ref{fig:automated-workflow}. All experiments were repeated at least five times at different times of the day to account for background activity variability. Results are reported as mean $\pm 1$ standard deviation. Each run proceeds through the following steps:

\begin{itemize}
    \item \textbf{Configuration.} Experimental parameters are parsed considering the scheduling approach (\texttt{CODECO} or default \texttt{K8s}), scenario
    (\texttt{S1}-\texttt{S5}), number of replicas, RPS intensity, and
   energy-awareness strategy (\texttt{green}, \texttt{comp}, or
    \texttt{net}). Then the corresponding strategy resource is deployed to the cluster.

    \item \textbf{Pre-clean.} All existing application workloads and ACM resources
    are removed to ensure a clean cluster state. \textit{DaemonSet} pods such as
    Kepler and k3s system services remain active throughout.

    \item \textbf{Workload deployment.} Either the CODECO-managed or default
    Kubernetes Bookinfo deployment is applied, and the script waits until all
    pods reach \texttt{Running} state before proceeding.

    \item \textbf{Baseline energy collection.} A pre-load snapshot of
    node-level energy metrics is collectd by Kepler.

    \item \textbf{Load generation and monitoring.} Apache JMeter generates
    foreground HTTP traffic at the configured RPS targeting the Bookinfo
    \texttt{productpage} microservice. Concurrently, a \texttt{LiveMonitor}
    records per-node energy and scheduling metrics every 60\,s into a
    timestamped CSV file. For scenarios with background interference
    (\texttt{S2}-\texttt{S5}), CPU stress and/or \texttt{iperf} network
    load is additionally injected via Chaos Mesh.

    \item \textbf{Post-load collection.} After a 600\,s monitoring window,
    post-load energy metrics are collected from Prometheus/Kepler.

    \item \textbf{Cleanup and cooldown.} All workloads and background stress
    processes are terminated, and the cluster is allowed a 180\,s cooldown
    before the next run begins.
\end{itemize}

To quantify the impact of the CODECO energy-aware scheduling strategy
relative to the plain Kubernetes scheduler, the energy increase index $ei$
is defined in Equation~\ref{eq:ei} as the relative increase in cluster energy
consumption from the pre-deployment baseline $e_{s1}$ to the post-deployment
steady state $e_{s2}$:
 
\begin{equation}
ei = \frac{e_{s2} - e_{s1}}{e_{s1}} \cdot 100
\label{eq:ei}
\end{equation}

\section{Performance Evaluation and Analysis}
\label{evaluation}
\subsection{Scenario~I: Scalability and Load Characterisation}

\begin{figure*}[htp!]
\centering
\subfloat[rep\,=\,1]{
    \includegraphics[width=0.8\textwidth]{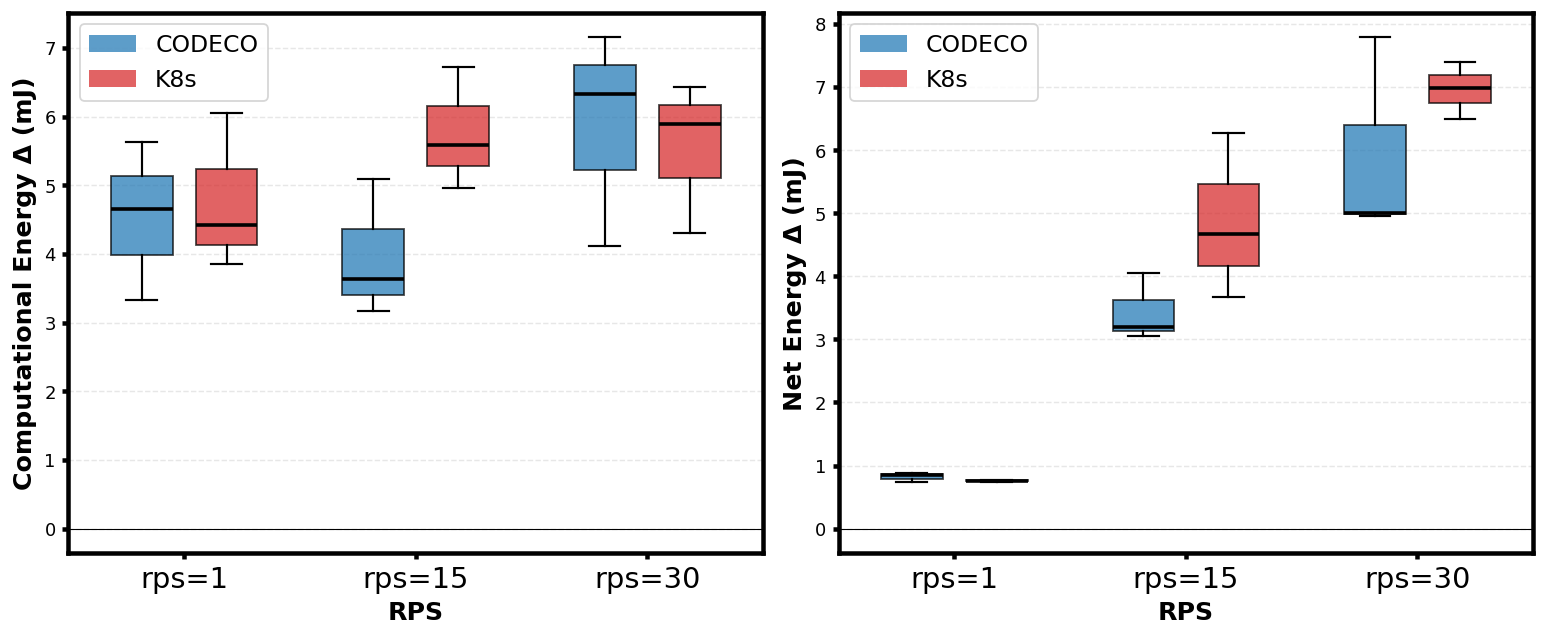}
    \label{fig:s1-box-rep1}
}
\\[6pt]
\subfloat[rep\,=\,4]{
    \includegraphics[width=0.8\textwidth]{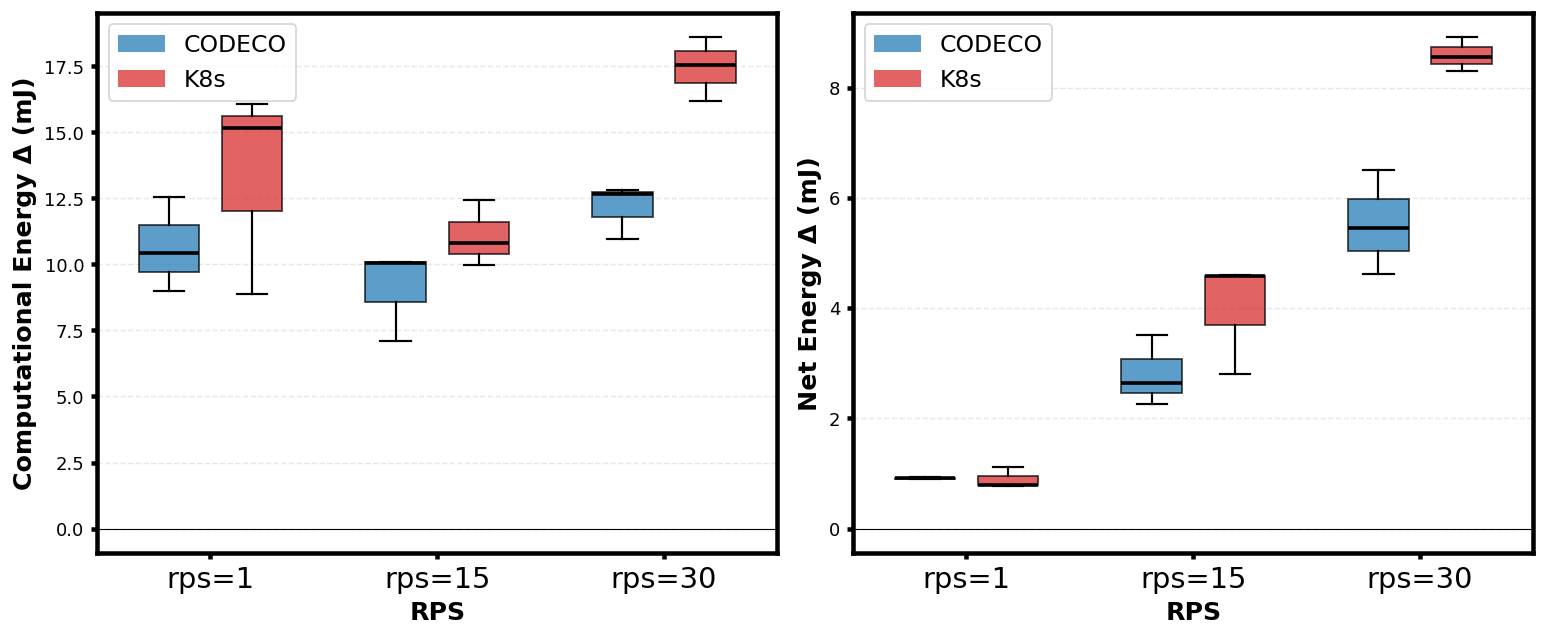}
    \label{fig:s1-box-rep4}
}
\\[6pt]
\subfloat[rep\,=\,7]{
    \includegraphics[width=0.8\textwidth]{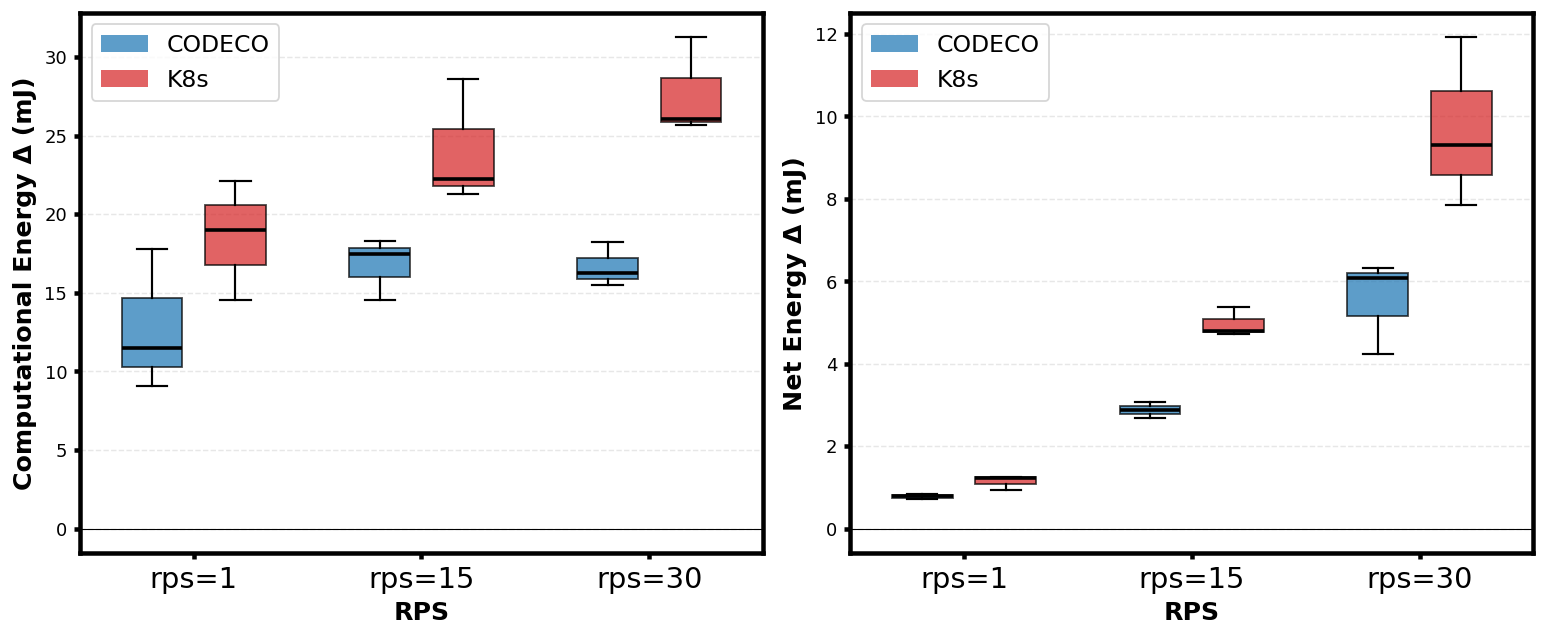}
    \label{fig:s1-box-rep7}
}
\caption{Computational and network energy boxplots across
         RPS\,=\,\{1,\,15,\,30\} for replica counts of
         1\,(a), 4\,(b), and 7\,(c).}
\label{fig:s1-boxplots}
\end{figure*}


\textbf{Scalability: impact of replica count.}
As shown in Figure~\ref{fig:s1-boxplots}, the energy behavior of the two frameworks diverges increasingly with replica count. Table \ref{tab:s1-results} provides a summary of the mean energy consumption gains across all use-cases.
\begin{table*}[htp!]
\centering
\caption{Average energy consumption $\pm$ standard deviation (mJ) and
         CODECO saving ($\Delta$ = K8s $-$ CODECO) across replica counts
         and RPS intensity.}
\label{tab:s1-results}
\setlength{\tabcolsep}{5pt}
\begin{tabular}{|c|c|c|c|c|c|c|c|}
\hline
\multirow{2}{*}{\textbf{Replicas}} &
\multirow{2}{*}{\textbf{RPS}} &
\multicolumn{3}{c|}{\textbf{Compute Energy (mJ)}} &
\multicolumn{3}{c|}{\textbf{Network Energy (mJ)}} \\
\cline{3-8}
& & \textbf{CODECO} & \textbf{K8s} & \textbf{Saving} &
      \textbf{CODECO} & \textbf{K8s} & \textbf{Saving} \\
\hline
\multirow{3}{*}{1}
 & 1  & $4.54 \pm 1.16$ & $4.78 \pm 1.14$ & \phantom{$-$}$0.24$ & $0.82 \pm 0.08$ & $0.76 \pm 0.02$ & $-0.06$ \\
 & 15 & $3.97 \pm 1.00$ & $5.76 \pm 0.89$ & \phantom{$-$}$1.79$ & $3.44 \pm 0.54$ & $4.87 \pm 1.31$ & \phantom{$-$}$1.43$ \\
 & 30 & $5.87 \pm 1.57$ & $5.55 \pm 1.10$ & $-0.32$             & $5.92 \pm 1.62$ & $6.96 \pm 0.45$ & \phantom{$-$}$1.04$ \\
\hline
\multirow{3}{*}{4}
 & 1  & $10.67 \pm 1.80$ & $13.37 \pm 3.91$ & \phantom{$-$}$2.70$ & $0.92 \pm 0.02$ & $0.89 \pm 0.19$ & $-0.03$ \\
 & 15 & $9.08  \pm 1.73$ & $11.08 \pm 1.25$ & \phantom{$-$}$2.00$ & $2.81 \pm 0.64$ & $4.00 \pm 1.03$ & \phantom{$-$}$1.19$ \\
 & 30 & $12.14 \pm 1.04$ & $17.43 \pm 1.21$ & \phantom{$-$}$5.29$ & $5.52 \pm 0.94$ & $8.59 \pm 0.31$ & \phantom{$-$}$3.07$ \\
\hline
\multirow{3}{*}{7}
 & 1  & $12.78 \pm 4.52$ & $18.54 \pm 3.83$ & \phantom{$-$}$5.76$ & $0.78 \pm 0.06$ & $1.14 \pm 0.18$ & \phantom{$-$}$0.36$ \\
 & 15 & $16.77 \pm 1.97$ & $24.04 \pm 3.96$ & \phantom{$-$}$7.27$ & $2.88 \pm 0.20$ & $4.97 \pm 0.36$ & \phantom{$-$}$2.09$ \\
 & 30 & $16.66 \pm 1.40$ & $27.67 \pm 3.11$ & \phantom{$-$}$11.01$& $5.55 \pm 1.14$ & $9.69 \pm 2.06$ & \phantom{$-$}$4.14$ \\
\hline
\end{tabular}
\end{table*}

With a single replica, CODECO and vanilla Kubernetes produce comparable computational and network energy across all RPS intensities. This is expected, given that a single replica offers no co-location opportunity and both schedulers make equivalent placement decisions in what concerns energy consumption. However, when the replica count increases, K8s exhibits higher and more variable computational energy,
particularly for the scenarios with lower RPS intensity, while network transmission energy diverges progressively as RPS increases. For instance, for 7 replicas, K8s scales steeply in computational energy with both replica count and RPS, whereas CODECO remains comparatively stable.
The same pattern of behavior occurs for the net strategy. 

The key difference is that CODECO scheduler relies on an ILP-based solver that provides a solution for an application and its micro-services. Hence, replicas are consolidated onto nodes that already host communicating peers. Fewer active nodes means lower aggregate idle and dynamic power. In contrast, K8s targets load-balancing and therefore schedules each pod independently, balancing CPU/memory requests across nodes, which tends to spread replicas and keep more nodes partially loaded and may result in a less energy-efficient operating point. Hence, for larger numbers of replicas, this becomes a disadvantage. For instance, for the case of seven replicas, they are possibly distributed across the target nodes to assure better load distribution, whereas CODECO consolidates them on a few nodes.

In regards to the network energy target (net), co-located replicas communicate via loopback interfaces rather than the Wi-Fi interface, eliminating the per-packet transmission energy entirely for intra-node flows. On the Wi-Fi testbed (IEEE 802.11bg), inter-node communication carries a non-negligible energy cost per packet. As K8s adds replicas across nodes, the volume of Wi-Fi traffic grows proportionally. CODECO's network-aware cost function explicitly penalizes link energy, so the scheduler actively avoids placements that generate inter-node flows.

\textbf{Load Characterisation: impact of RPS.}
RPS is the primary driver of network energy differences between the two frameworks, and this can be better observed in Table ~\ref{tab:s1-results}. Across all replica counts, the gap between CODECO and K8s widens as the request intensity increases, and this effect is most visible for a larger number of replicas (rf. to Figure 7(c)). An increase in RPS increases the packet rates of active IP flows (more packets per second, higher dynamic and static energy costs), hence directly increasing link energy for the linear power model provided in Equation ~\ref{eq:link_energy}.

Computational energy also rises with an increase in the RPS intensity. However, the per-request CPU work (HTML rendering, database lookup in Bookinfo) is small and does not scale with request rate in a way that materially increases node power draw. The combination of high RPS and high replica count is therefore a worst-case for K8s (and a best case for CODECO), as it simultaneously maximises both the number of inter-node flows and their packet rate .



\begin{figure*}[t]
  \centering
  \subfloat[Cluster Energy]{
    \includegraphics[width=\textwidth]{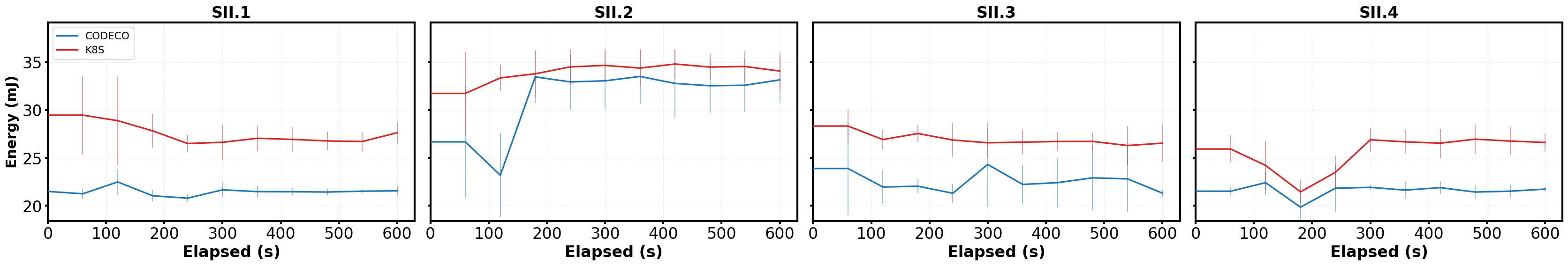}
    \label{fig:cluster_energy}
  }
  \\[4pt]
  \subfloat[Cluster Net Energy]{
    \includegraphics[width=\textwidth]{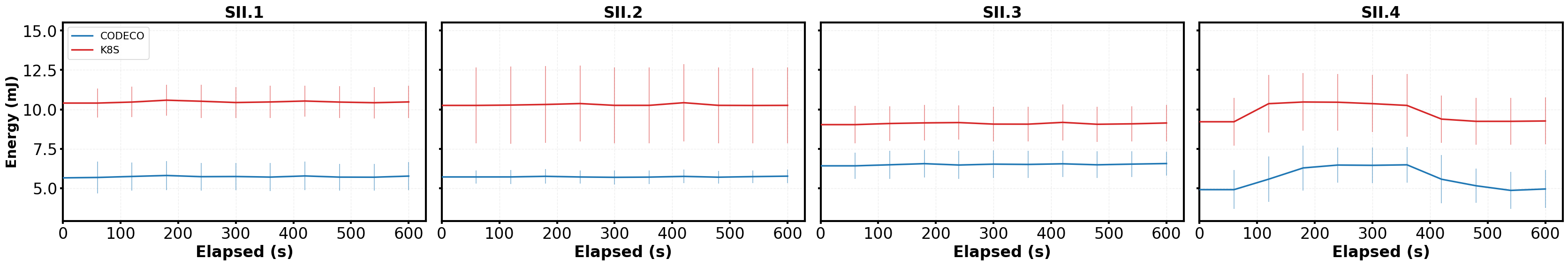}
    \label{fig:cluster_net_energy}
  }
  \caption{Time-series evolution of cluster compute energy (top) and
network energy (bottom) across sub-scenarios
SII.1--SII.4 (rep\,=\,7, RPS\,=\,30).}
  \label{fig:sII-energy_timeseries}
\end{figure*}

\subsection{Scenario~II: Edge Fault Resilience}
Figure~\ref{fig:sII-energy_timeseries} presents the time-series evolution of
cluster energy and network transmission energy across sub-scenarios
SII.1--SII.4 at the operating point with rep equal to 7 and RPS equal to 30.
Each chart reports mean values across repeated runs with one standard
deviation bands. Experiments were run for 600\,s.

\textbf{SII.1 Baseline.}
Under nominal conditions, CODECO maintains a stable compute cluster energy of approximately 21\,mJ throughout the 600\,s window, compared with
approximately 29\,mJ for K8s. This indicates that CODECO consistently requires less compute energy while exhibiting a similarly stable temporal
behavior. Network energy is also stable: CODECO remains at approximately 5.5\,mJ, whereas K8s stays around approximately 10\,mJ. This baseline
establishes the reference separation attributable solely to placement strategy under undisturbed conditions.

\textbf{SII.2 CPU Stress Impact.}
Under \texttt{StressChaos} faults, CODECO exhibits a pronounced dip-and-recovery pattern in compute cluster energy. This behavior appears
to be related to the following operation. Upon detecting CPU stress,
workloads are migrated to less-loaded nodes, causing Kepler telemetry to
temporarily underestimate cluster consumption as pods transition. This
produces the sharp drop to approximately 22\,mJ near 100\,s. Energy then
rises toward the post-migration steady state and converges close to K8s,
with both systems operating at approximately 34\,mJ. The wide error bands
that CODECO exhibits during the early migration window reflect run-to-run
variability in migration timing. Hence, the initial compute-energy advantage
is significantly reduced under CPU stress once migration completes, as both
K8s and CODECO possibly operate on similarly loaded nodes. For network
energy, CODECO remains consistently lower than K8s, suggesting that
co-location is preserved across the migration event and that the
transmission-efficiency gap is maintained.

\textbf{SII.3 Network Delay Impact.}
Under \texttt{NetworkChaos} latency injection, both time-series remain flat.
With CODECO, the resulting cluster compute energy remains at approximately
23\,mJ, compared with approximately 28\,mJ for K8s. CODECO also maintains
lower cluster network energy, at approximately 6.5\,mJ against approximately
9\,mJ for K8s. However, unlike the baseline case in SII.1, the
network-energy gap does not widen; instead, it narrows from approximately
4.5\,mJ to approximately 2.5\,mJ. This indicates that latency injection does
not amplify K8s network-energy overhead in this experiment. Rather, CODECO
remains more energy-efficient, although its relative network-energy
advantage is reduced under the delay-injection condition.

\textbf{SII.4 Controlled Bandwidth Contention Impact.}
Background \texttt{iperf3} traffic introduces visible transient behavior in
both metrics. Compute energy dips briefly in both frameworks near 150\,s
before recovering. After recovery, CODECO stabilises at approximately
22\,mJ, while K8s stabilises at approximately 27\,mJ. Network transmission
energy rises transiently for CODECO to approximately 7.5\,mJ during the
congestion window before recovering toward its baseline level. In contrast,
K8s remains elevated at approximately 10\,mJ without a comparable recovery.

\textbf{SII Cross-scenario summary.}
Across all four sub-scenarios, CODECO consistently results in lower compute cluster energy and network energy than K8s. Three patterns are observable in
Figure~\ref{fig:sII-energy_timeseries}. First, an  improvement in regards to energy consumption due to computational resources is
most stable and pronounced for the baseline case in SII.1 and for the network-delay case in SII.3, reflecting the effect of undisturbed placement
decisions. Second, CPU stress in SII.2 temporarily narrows this advantage as CODECO migrates workloads. After migration, comp energy consumption
moves closer to the K8s level while still remaining lower. 

As for the improvement concerning energy consumption due to packet transmission, it is stable throughout the migration event, indicating that co-location
survives the fault-handling process. SII.4 is the only sub-scenario where CODECO's network transmission energy rises transiently above its baseline, yet it recovers while K8s remains elevated. This suggests that the CODECO energy-aware placement strategy is more
resilient to congestion than K8s.

\begin{figure*}[htp!]
\centering
\includegraphics[width=\textwidth]{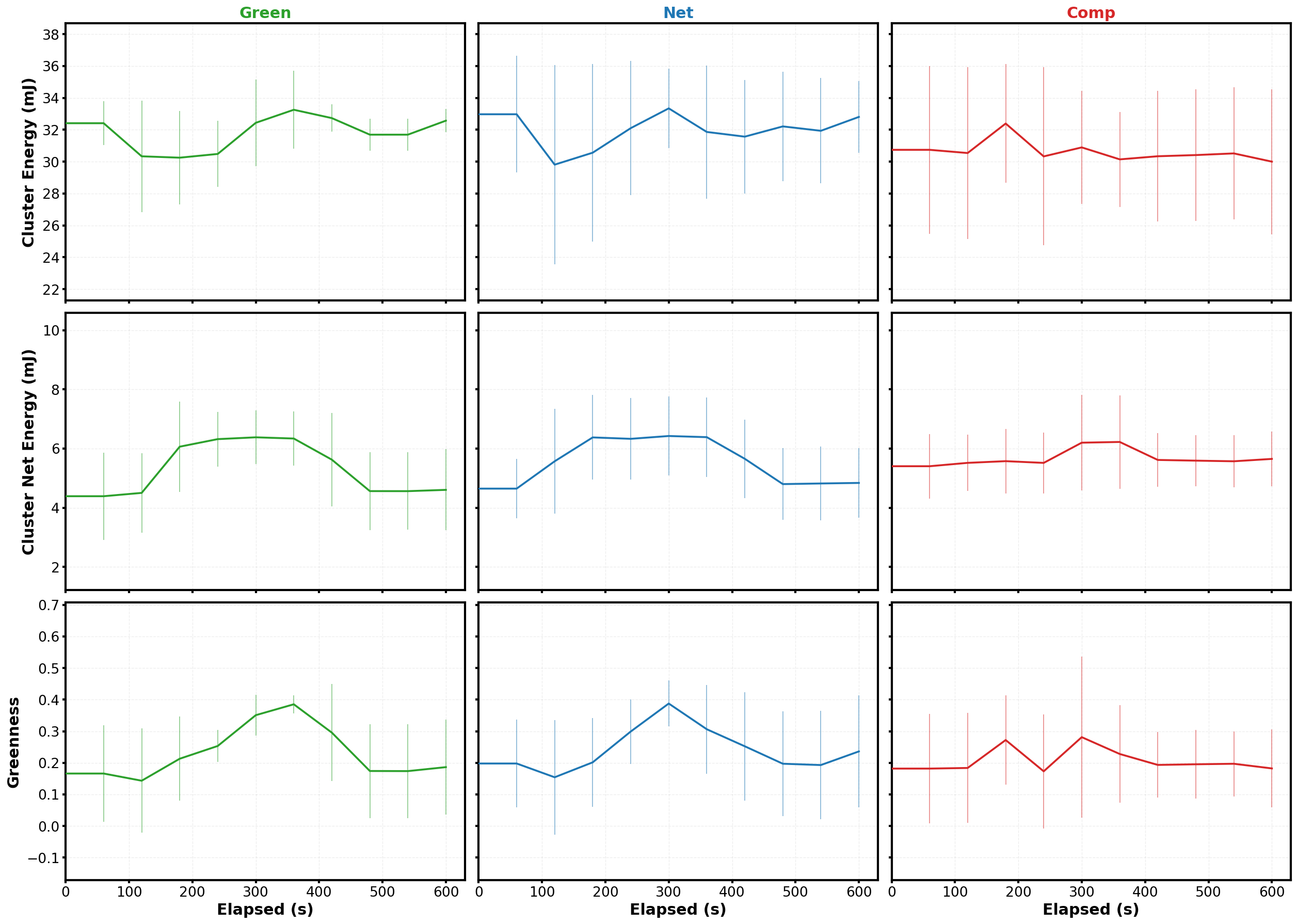}
\caption{Time-series evolution of cluster compute energy (top row), network energy (middle row), and greenness (bottom row) under combined CPU and bandwidth stress (S5) for the three CODECO scheduling objectives: \texttt{green}, \texttt{net} and \texttt{comp}.}
\label{fig:s3-strategy}
\end{figure*}
\subsection{Scenario~III: Greenness Scheduling Strategies Comparison}

Figure~\ref{fig:s3-strategy} presents the time-series evolution of
cluster energy, network energy, and composite greenness scores for the
three $g(i)$ scheduling cost functions, under combined CPU and bandwidth stress.

\textbf{Computational energy.}
All three $g(i)$ functions produce broadly comparable cluster energy
consumption throughout the experiments, with most values remaining around
approximately 31\,mJ. \texttt{comp} provides the lowest and most stable
profile, staying close to approximately 30\,mJ, which is consistent with
its direct optimisation of computational cost. \texttt{net} exhibits a
transient dip near 100\,s before recovering to approximately 33\,mJ, while
\texttt{green} fluctuates more visibly, reaching its lowest point at
approximately 30\,mJ and increasing to approximately 33\,mJ near 350\,s.
The narrow spread across functions indicates that cluster-level
computational energy is not strongly discriminated by the choice of
$g(i)$ under combined stress.

\textbf{Network energy.}
Clearer differentiation emerges in network transmission energy consumption.
\texttt{green} exhibits the most pronounced rise during the stress window,
peaking at approximately 8\,mJ before declining to approximately 6.5\,mJ
after 500\,s. \texttt{net}, which explicitly targets transmission energy,
rises transiently to approximately 7.3\,mJ around 300\,s but recovers more
sharply, converging to approximately 5.8\,mJ, the lowest post-stress value
among the three functions. \texttt{comp} shows the flattest trajectory
throughout the experiment, remaining close to approximately 6.5\,mJ.
Although it does not explicitly target network transmission, consolidating
replicas for computational efficiency incidentally reduces inter-node
flows, thereby stabilising transmission energy under combined fault
conditions.

\textbf{Composite greenness as a ranking indicator.}
The composite greenness score, defined as in Equation \ref{eq:greenness_composite}, consolidates
both energy dimensions into a single indicator and is evaluated here as a
tool for ranking the three $g(i)$ functions. The score produces a
consistent and stable ordering throughout the full 600\,s window.
\texttt{comp} achieves the lowest score, staying around approximately
0.15. \texttt{net} occupies the intermediate position, peaking at
approximately 0.32 before recovering to approximately 0.15. \texttt{green}
scores highest, peaking at approximately 0.38 before declining to
approximately 0.20.

The ordering reflects how each $g(i)$ function affects the two constituent
dimensions of the score. \texttt{comp} captures computational energy directly and, as a side effect, co-locates replicas. This reduces inter-node flows and keeps network energy low, suppressing both dimensions of the score simultaneously and yielding the lowest
composite value. \texttt{net} targets network
transmission energy explicitly but does not constrain computational placement, leaving cluster energy higher and more variable.  \texttt{green} jointly targets both dimensions. However, under simultaneous CPU and network stress, it receives conflicting cost signals from the two terms, preventing it from suppressing either dimension effectively. The network transmission term dominates the score at peak stress, producing the highest composite value despite the composite intent
of the function.

The experiments run show that the composite greenness score is sensitive enough to discriminate the three scheduling functions and produce a stable cluster ranking under stress. However, the result also exposes a
limitation of the composite $g(i)$ formulation: jointly targeting both dimensions does not guarantee a joint energy advantage under concurrent fault conditions. Refinement, for instance through adaptive weighting of
the constituent terms based on the observed fault type, is needed before \texttt{green} can be reliably used as a scheduling objective in degraded
edge environments.

\subsection{Analysis Summary and Discussion}
The three scenarios collectively address three questions: i) whether CODECO reduces energy consumption relative to K8s; ii) which $g(i)$ scheduling function performs best; and iii) whether the composite greenness
score is an effective indicator for cluster ranking.

\textbf{Is CODECO more energy-efficient than K8s?}
The results consistently support this claim. First, the advantage grows with replica count. This reflects CODECO's application-first deployment model: the ILP solver places the full microservice graph jointly, prioritizing co-location over load balancing across nodes. Second, the two energy dimensions respond to different drivers. Network energy savings are governed primarily by RPS intensity as higher traffic amplifies the cost of inter-node flows generated by K8s's distributed placement. Computational energy savings are governed primarily by replica count as more replicas mean more nodes partially
loaded under K8s, while CODECO consolidates them onto fewer nodes.
These two dimensions are therefore partially independent: a single-dimension $g(i)$ function may leave savings in the other dimension unrealised, reinforcing the need to monitor both metrics separately when assessing overall energy efficiency.

Under fault conditions (Scenario~II), the computational advantage narrows temporarily during workload migration but recovers, and the network energy gap is preserved across all sub-scenarios. This confirms that CODECO's energy efficiency is robust to the injected fault conditions evaluated, including CPU stress, asymmetric network
delay, and controlled bandwidth contention.

\textbf{Is there a best $g(i)$ function?}
No single $g(i)$ formulation dominates unconditionally across all conditions, but a clear ordering emerges under combined stress.
\texttt{comp} achieves the best outcome across all three cluster-level metrics: lowest computational energy, flattest network energy trajectory, and lowest composite greenness score. This is counterintuitive: a compute-only function outperforms the composite formulation on the composite metric. The reason is structural: co-locating replicas for computational efficiency simultaneously eliminates inter-node flows, suppressing network energy as a side effect. Both dimensions of the greenness score are therefore reduced without the $g(i)$ function explicitly targeting either
jointly.

\texttt{net} achieves the fastest post-stress recovery in network transmission energy ($\approx$5.8\,mJ) and is preferable when rapid restoration of network efficiency after a fault is the operational priority. \texttt{green} performs worst during the fault period across all metrics: its composite cost function receives conflicting signals from both dimensions simultaneously under combined stress, preventing it from suppressing either effectively. In its current form it is least suited to degraded edge environments. It remains a candidate for stable, lightly loaded conditions where both cost signals are unambiguous and the joint optimisation does not face conflicting pressure.

The choice of $g(i)$ should therefore be guided by the operational context: \texttt{comp} for sustained energy minimization under stress, \texttt{net} for fast post-fault network recovery, and \texttt{green} for undisturbed conditions or as a starting point for further refinement, for instance, through adaptive weighting of its
constituent terms based on observed infrastructure state.

\textbf{Is the composite greenness score an effective ranking
indicator?}
The composite greenness score maintains a consistent ordering of
the three $g(i)$ functions across all experimentation, including during fault windows. This stability demonstrates that the score captures the joint energy behavior of each scheduling function and produces a reliable cluster ranking under stress. It is therefore more informative than either single-dimension metric alone when comparing
scheduling strategies. However, when diagnosing the cause of a ranking outcome, the two constituent scores must be inspected separately: the composite value masks whether a high score is driven by computational overhead, network overhead, or both.

\textbf{Limitations.}
Three additional aspects are worth noting. First, workload migration
carries a non-negligible energy cost: the dip-and-recovery pattern
observed under CPU stress temporarily converges CODECO's cluster
energy toward K8s levels. Frequent rescheduling in highly dynamic
environments could erode the steady-state energy advantage, and
migration-aware scheduling strategies that explicitly account for
relocation overhead are an important direction for future work.
Second, measurement fidelity on ARM-based embedded devices is
limited: Kepler's power model was trained on x86 hardware and
provides only approximate estimates on Raspberry Pi~4 nodes,
introducing a systematic uncertainty that affects all results equally
but limits the precision of absolute energy values. Third, the Wi-Fi
connectivity (IEEE 802.11bg) amplifies network energy differences
relative to wired deployments; the advantage of co-location in
eliminating inter-node transmission energy may be less pronounced in
environments with lower per-packet transmission costs.

\section{Conclusions and Next Steps}
\label{conclusions}

This work evaluates the energy efficiency of the K8s CODECO orchestration
framework in a single-cluster far Edge deployment, focusing on how
scheduling decisions affect both computational and network
energy consumption. Three experimental scenarios are evaluated on a real-world
ARM-based testbed, comparing CODECO against vanilla K8s across
varying workload intensities, injected fault conditions, and
alternative energy-aware scheduling functions.

The results consistently demonstrate that CODECO reduces cluster
energy consumption relative to vanilla K8s. The advantage is
driven by two independent mechanisms: CODECO's ILP-based batch
scheduler gives priority to an application dimension, thus co-locating communicating micro-service replicas, reducing the number of active nodes and eliminating inter-node Wi-Fi
transmission for intra-application flows; and its reactive stateful migration mechanism reallocates workloads away from degraded nodes under fault conditions, preserving energy efficiency across CPU stress, asymmetric network delay, and bandwidth contention. The benefit scales with replica count for computational energy, and with request rate for
network  energy, and is most pronounced at their
combination.

Regarding the $g(i)$ scheduling functions, no single formulation
reaches a best target unconditionally. Under combined stress, the compute-only
function (comp) achieves the best outcome across all three cluster-level metrics, including the composite greenness score, because co-locating for computational efficiency
incidentally suppresses inter-node traffic as a side effect. The network function (net) recovers most rapidly in transmission energy after fault removal. The composite function (green) performs worst under combined stress, as simultaneous CPU and network degradation generates conflicting cost signals that prevent either dimension from being
suppressed effectively. 
Finally, the composite greenness score proposed provides a stable and consistent ranking of the three functions throughout all experimental conditions, making it a reliable indicator for cluster-level energy assessment.

These findings point to three directions for future work.

\textbf{Migration-aware scheduling:} workload relocation carries a non-negligible energy cost that the current policy does not account for, and strategies that explicitly model and minimise the overhead of pod migration are needed for highly dynamic edge environments.

\textbf{Architecture-specific energy modeling: }Kepler's power model was trained on x86 hardware and provides only approximate estimates on ARM-based embedded devices; retraining or calibrating the model against the target hardware would improve the fidelity of energy
measurements and the accuracy of $g(i)$ cost computations. 

\textbf{Refinement of the $g(i)$ scheduling heuristics:} the experiments confirm that the composite formulation is the most sensitive of the three scoring functions to concurrent fault conditions, as it responds to cost signals from both energy dimensions simultaneously. Under combined CPU and network stress this sensitivity produces higher variability in the greenness score, as neither dimension is fully suppressed when both are degraded at once. This same sensitivity, however, makes the composite $g(i)$ the most promising candidate for refinement: a formulation that weights the constituent terms adaptively based on observed infrastructure state increasing the network weight under congestion, the computational weight under CPU stress, and
rebalancing under combined conditions could make \texttt{green}
the most responsive and accurate scheduling function across all operating conditions, rather than only under undisturbed operation. A further refinement direction is to incorporate migration cost explicitly into the $g(i)$ formulation, so that the scheduler
accounts for the energy overhead of workload relocation before triggering a migration, addressing the transient inefficiency observed under CPU stress in Scenario~II.
\textbf{Security-aware greenness scheduling: }The experiments presented in this paper are conducted on a controlled, isolated laboratory network and do not incorporate communication security overhead into the scheduling model. In real-world CEI deployments, encryption, mutual authentication, and secure overlay management — capabilities provided by the CODECO NetMA secure connectivity sub-component introduce non-trivial computational and transmission energy costs that vary with traffic volume and cipher suite. Extending the g(i) scheduling heuristics to incorporate security-induced energy as an additional cost dimension is an open problem and a natural direction for future work, particularly in IIoT settings where both sustainability and confidentiality are operational requirements.

\section*{Acknowledgement}
This work has been funded by The European Commission in the context of the Horizon Europe CODECO project under grant number 101092696, and by SGC, Grant agreement nr: M-0626, project SemComIIoT.

\def\refname{REFERENCES}
  \bibliographystyle{ieeetr}
  \bibliography{references}

@misc{sofia2026scalablefederatedcontainerorchestration,
      title={Towards Scalable Federated Container Orchestration: The CODECO Approach}, 
      author={Rute C. Sofia and Josh Salomon and Ray Carrol and Luis Garcés-Erice and Peter Urbanetz and Jürgen Gesswein and Rizkallah Touma and Alejandro Espinosa and Luis M. Contreras and Vasileios Theodorou and George Papathanail and Georgios Koukis and Vassilis Tsaoussidis and Alberto del Rio and David Jimenez and Efterpi Paraskevoulakou and Panagiotis Karamolegkos and John Soldatos and Borja Dorado Nogales and Alejandro Tjaarda},
      year={2026},
      archivePrefix={arXiv},
      primaryClass={cs.DC},
      url={https://arxiv.org/abs/2601.13351}, 
}

@inproceedings{feeney2001,
  author       = {Feeney, L. M. and Nilsson, M.},
  booktitle    = {Proceedings IEEE INFOCOM 2001. Conference on
                  Computer Communications. Twentieth Annual Joint
                  Conference of the IEEE Computer and Communications
                  Society},
  title        = {Investigating the energy consumption of a wireless
                  network interface in an ad hoc networking
                  environment},
  year         = {2001},
  volume       = {3},
  pages        = {1548--1557},
  doi          = {10.1109/INFCOM.2001.916651},
}

@misc{kepler,
  title = {Kepler: A Sustainable Computing Framework},
  author = {{Sustainable Computing}},
  year = {2023},
  url = {https://sustainable-computing.io/design/metrics/},
  note = {Accessed: 2024-10-22}
}

@misc{CODECO-D31,
  author       = {{Sofia, Rute C. (ed.)}},
  title        = {{D31: CODECO} Federated Cluster Operation
                  Architectural Design},
  month        = may,
  year         = {2026},
  publisher    = {Zenodo},
  doi          = {10.5281/zenodo.20049338},
  url          = {https://doi.org/10.5281/zenodo.20049338},
}

@ARTICLE{sofia2024framework,
  author={Sofia, Rute C. and Salomon, Josh and Ferlin-Reiter, Simone and Garcés-Erice, Luis and Urbanetz, Peter and Mueller, Harald and Touma, Rizkallah and Espinosa, Alejandro and Contreras, Luis M. and Theodorou, Vasileios and Psaromanolakis, Nikos and Mamatas, Lefteris and Tsaoussidis, Vassilis and Fu, Xiaoming and Yuan, Tingting and del Rio, Alberto and Jiménez, David and Stam, Andries and Paraskevoulakou, Efterpi and Karamolegkos, Panagiotis and Vieira, Vitor and Martrat, Josep and Prusiel, Ignacio Mariscal and Matzakou, Dorine and Soldatos, John and Remon, David and Jahn, Marco},
  journal={IEEE Access}, 
  title={A Framework for Cognitive, Decentralized Container Orchestration}, 
  year={2024},
  volume={12},
  number={},
  pages={79978-80008},
   url={https://ieeexplore.ieee.org/document/10540390},
  doi={10.1109/ACCESS.2024.3406861}}

@misc{Pol2024,
  author       = {Pol, Ties},
  title        = {Carbon Footprint Monitoring up to Container-Level in Virtualized Environments: A Hardware and Hypervisor-Free Approach},
  year         = {2024},
  type         = {Master's Thesis / Essay},
  institution  = {University of Groningen},
  degree       = {Computing Science},
  supervisor   = {Andrikopoulos, V. and Setz, B.},
  url          = {https://fse.studenttheses.ub.rug.nl/id/eprint/32271},
  note         = {Accessed: 2024-04-12}
}

@article{luo2019fog,
title = {Container-based fog computing architecture and energy-balancing scheduling algorithm for energy IoT},
journal = {Future Generation Computer Systems},
volume = {97},
pages = {50-60},
year = {2019},
issn = {0167-739X},
doi = {https://doi.org/10.1016/j.future.2018.12.063},
url = {https://www.sciencedirect.com/science/article/pii/S0167739X1831358X},
author = {Juan Luo and Luxiu Yin and Jinyu Hu and Chun Wang and Xuan Liu and Xin Fan and Haibo Luo}
}

@INPROCEEDINGS{netmarks,
  author={Wojciechowski, {\L}ukasz and Opasiak, Krzysztof and Latusek, Jakub and Wereski, Maciej and Morales, Victor and Kim, Taewan and Hong, Moonki},
  booktitle={IEEE INFOCOM 2021 - IEEE Conference on Computer Communications}, 
  title={NetMARKS: Network Metrics-AwaRe Kubernetes Scheduler Powered by Service Mesh}, 
  year={2021},
  volume={},
  number={},
  pages={1-9},
  doi={10.1109/INFOCOM42981.2021.9488670}}

@ARTICLE{zeus,
  author={Zhang, Xiaolong and Li, Lanqing and Wang, Yuan and Chen, E. and Shou, Lidan},
  journal={IEEE Access}, 
  title={Zeus: Improving Resource Efficiency via Workload Colocation for Massive Kubernetes Clusters}, 
  year={2021},
  volume={9},
  number={},
  pages={105192-105204},
  doi={10.1109/ACCESS.2021.3100082}}

@INPROCEEDINGS{nas,
  author={{Ragavan, Vivek Karunai Kiri and Nadig, Deepak}},
  booktitle={ICC 2025 - IEEE International Conference on Communications}, 
  title={NAS: A Novel Network-Aware Kubernetes Scheduling Framework Using eBPF Service Mesh}, 
  year={2025},
  volume={},
  number={},
  pages={1512-1517},
  doi={10.1109/ICC52391.2025.11161235}}

@misc{codeco_energy,
      title={Experimenting with Energy-Awareness in Edge-Cloud Containerized Application Orchestration}, 
      author={Dalal Ali and Rute C. Sofia},
      year={2025},
      eprint={2511.09116},
      archivePrefix={arXiv},
      primaryClass={cs.NI},
      url={https://arxiv.org/abs/2511.09116}, 
}

@inproceedings{RPIs,
  author       = {Ardito, Luca and Torchiano, Marco},
  title        = {Creating and evaluating a software power model for
                  {Linux} single board computers},
  booktitle    = {Proceedings of the 6th International Workshop on
                  Green and Sustainable Software},
  series       = {GREENS '18},
  year         = {2018},
  pages        = {1--8},
  publisher    = {Association for Computing Machinery},
  address      = {New York, NY, USA},
  doi          = {10.1145/3194078.3194079},
  url          = {https://doi.org/10.1145/3194078.3194079},
}

@misc{cncf2023,
  author       = {{Cloud Native Computing Foundation}},
  title        = {Exploring {Kepler}'s potentials: Unveiling cloud
                  application power consumption},
  year         = {2023},
  month        = oct,
  url          = {https://www.cncf.io/blog/2023/10/11/exploring-keplers-potentials-unveiling-cloud-application-power-consumption/},
  note         = {Accessed: 2024-11-13},
}
\end{document}